\documentclass[preprint,12pt]{elsarticle}

\usepackage{graphicx}

\usepackage{amssymb}
\usepackage{epsf,colordvi,color,amsbsy}

\begin{document}

\begin{frontmatter}

\title{Ball lightning observation: an objective video-camera analysis report}

\author[label1]{Stefano Sello \corref{cor1}},
\author[label2]{Paolo Viviani},
\author[label1]{Enrico Paganini}

\cortext[cor1]{stefano.sello@enel.it}

\address[label1]{Enel Research, Pisa - Italy}
\address[label2]{Business-e an Itway Group Company, Massa - Italy}

\begin{abstract}
In this paper we describe a video-camera recording of a (probable) ball lightning event and both the related image and signal analyses for its photometric and dynamical characterization. The results strongly support the BL nature of the recorded luminous ball object and allow the researchers to have an objective and unique video document of a possible BL event for further analyses. Some general evaluations of the obtained results considering the proposed ball lightning models conclude the paper.
\end{abstract}
\end{frontmatter}

\section{Introduction}
In this section we review all the basic properties of ball lightning as extensively reported in the scientific literature with an account of the main models proposed to explain some of their most peculiar properties. 

\subsection{General}
Two of the main characteristic features of ball lightning (in short BL) are the unpredictability of its behavior (formation, stability, motion, etc.) and the variability of its properties (structure, size, color, temperature, etc.). Nevertheless, there are many sufficient qualitative similarities in the qualitative accepted properties to imply that ball lightning either is a real unique phenomenon or at least represents a homogeneous class of related physical phenomena.
Over the past 200 years, more than 2000 observations of ball lightning have been reported in sufficient details for scientists to take them seriously. They are almost invariably associated with stormy weather. Sometimes they just disappear, so they probably be gaseous. But it is not obvious how there can be a surface between two gases that is sufficiently stable to allow the BL to bounce or to squeeze through small holes. A possible explanation is provided by an approximate thermodynamic analysis of the process which must be occurring as ions escape from a hot air plasma into moist, electrically charged air. The process is the hydration of ions at high relative humidity which is basically an electrostatic phenomenon (Turner, 1994). 
Mainly after the books of Singer \cite{Singer71}, Stakhanov \cite{Stakhanov79}, Barry \cite{Barry80} and Stenhoff \cite{Stenhoff10} there is a clear evidence for the existence of BL as a real physical phenomenon. All books consider individual accounts and statistical analyses of a large number of observations. Another survey by Smirnov \cite{Smirnov87} contains an even larger collection of statistical information. Now there is a remarkable consensus about the main characteristics derived from over 2000 verified observations. BL are generally observed during thundery weather, though not necessarily during a storm. It appears to be a free floating globe of glowing gas, usually spherical in shape, which can enter buildings or airplanes. The formation of BL is rarely seen, although it has occasionally been observed forming from linear lightning in the sky and growing out of and detaching from electrical discharges on the ground. It has also been observed to fall from the cloud base. Rarely also it has been  reported BL rolling on or bouncing off, usually wet, surfaces. The credibility of the reports is an important issue when we handle with rare observations. It is apparent, on the basis of thousand of observations that the BL objects have considerable stability. Because they are gaseous this is a surprising feature and it has not yet received a satisfactory explanation.
There is a general qualitative agreement about most of the main characteristics of a BL, such as size, shape, colour, stability, temperature, liftime and demise. However some of the estimated ranges for the quantitative properties differ for different authors. For example, the estimated energy density is in the range $1-10 ~ J cm ^{-3}$ by Stakhanov \cite{Stakhanov79} or a wider range but below the value $0.2 ~ kJ cm ^{-3}$ by Barry \cite{Barry80}. Bichkov et al. \cite{Bichkov02} consider the most energetic BL effects and estimate a very high energy density $> 1 ~ kJ cm ^{-3}$ and this can be explained by a polymer composite model. The energy source for BL are generally divided into two basic groups: one assumes that the ball is continuously powered by some external sources, such as the electric field (ef) from clouds or a radio-frequency (rf) field transmitted from discharges in the clouds. The other group assumes that no external sources are needed and that the ball is generated with sufficient energy to sustain it for its full lifetime. This internal source of energy are mainly organic materials, unstable molecules and plasma from a lightning stroke from clouds.
A small number of the observations has been accurately investigated to determine the reliability of the eyewitness and to evaluate the related reports. None of the attempts has succeeded in obtaining even a photograph of the elusive phenomenon. In fact, only a few photographs have been obtained by chance by observers who also saw the object photographed, (Singer, \cite{Singer02}). A number of photographs of alleged ball lightning are of exceptionally poor quality and have insufficient detail for evaluation or to yield useful data (Stenhoff,\cite{Stenhoff10}). Reviews by Barry \cite{Barry80} and Stenhoff \cite{Stenhoff10} confirm that "the evidence provided by still photographs alleged to be of ball lightning is very questionable. Still photographs taken by chance will always be a matter of controversy. The likelihood of obtaining probative photographic evidence of ball lightning through chance observation is small. Videotapes (or films) have the potential to yield more useful data, but there is still the possibility of error." Indeed, the reported films contain artifacts of different nature (Stenhoff, \cite{Stenhoff10}).

\subsection{Size}

The sizes, as reported by observations, vary usually in the range 5-30 cm of diameter. The average observed diameter is 24 cm, but twice as large diameters are often observed, Birbrair \cite{Birbrair01}. It is important to note that the size of a typical BL depends on too many unknown quantities to allow a realistic prediction, Turner \cite{Turner94}. A characteristic almost always noted is that the size of the ball hardly changes during the observed lifetime.

\subsection{Temperature}
There is a wide range of plasma temperatures required by different analyses. One of the highest estimates for a BL was that of Dmitriev \cite{Dmitriev67}, \cite{Dmitriev69} who suggested a value of 14000 C. In the model of Powell and Finkelstein \cite{Powell69} the plasma temperature estimated by the radiation emitted with a radio-frequency excitation was 2000-2500 K.    
The core of a BL are hot enough to melt holes in glass. Very little heat appears to be emitted from the external surface. In the view of Stakhanov, with an internal source of energy, the BL plasma is of quite low temperature (500-700 K) with ions extensively hydrated. This low temperature is based mainly on heat loss calculations and on common evidence for the low surface temperature of some BL as reported by many direct  observations. Turner \cite{Turner94} considers more convincing all the evidence for a central plasma zone with a temperature of at least 2000 K.

\subsection{Color and brithness}

Color and brightness vary. The observed BL colors cover the region from $\lambda=7 \cdot 10 ^{-5} cm$ (red) to $\lambda=3.8 \cdot 10 ^{-5} cm$ (violet), Ofuruton et  al. \cite{Ofuruton97}. The fact that a wide range of colors has been reported seems to reflect the presence of impurities in the plasma and does not appear correlated with the size or other properties. The most common reported color is flame-like, approximatively orange, but occasionally brilliant white or red, blue or less often green (Singer, \cite{Singer02}).

\subsection{Dynamics}

Observed motion of BL: it frequently moves horizontally at speeds between 0.1 and 10 m/s and a meter or so above the ground. There are also reports of vertical motions or more irregular type. Following Stakhanov \cite{Stakhanov79}, we find that in 30\% of observations a slow rotation of the ball was reported.   
Mostly of BL lasts (or is observed) for less than 50 s, although some Russiam surveys reported that a BL can lasts over 100 s (Stakhanov, \cite{Stakhanov79}). During its life time it rarely changes significantly in either size or color but its life can end in two quite different ways: explosively or by simply disappearing. The motion of the ball, which is sometimes directly down from the clouds, sometimes upward from its appearance near the ground, sometimes in a straight line at low velocity, (Singer, \cite{Singer02}).

\subsection{Models}

The incompatibility of existing models with all the accepted qualitative properties of BL has led to many authors to propose and to prefer quite different models (e.g. Singer, \cite{Singer71}; Stakhanov, \cite{Stakhanov79}; Barry, \cite{Barry80}; Turner, \cite{Turner94}, \cite{Turner02}; Stenhoff, \cite{Stenhoff10}, etc.). First of all, it is not wide accepted that ball lightning is of one basic type phenomenon. The external source of energy can explains the long life, greater than 100 s, of some BL.  However, the internal source of energy better explains the frequently observed motion of the ball and its occasional appearance inside enclosed and electrically shielded areas. Singer \cite{Singer71} prefers an external supply of energy by rf induced powering, whereas Stakhanov \cite{Stakhanov79}, \cite{Stakhanov84} prefers an energy self-sufficient ball where the energy originates from the plasma of a near lightning stroke. Turner \cite{Turner94} considers the possibility of both internal and external sources of energy. In fact, the ball could store the electrical (or electromagnetic) energy received from a local discharge and use it to replace the external source when it becomes unavailable. In this case, the electric field of a thunderstorm provides most of the required external power, at least during the formation of a BL. Ignatovich \cite{Ignatovich06} considers an electromagnetic model i.e. a thin spherical layer filled with electromagnetic radiation retained because of the total internal reflection and the layer itself is conserved because of electrostiction forces generated by the radiation. This model can explain both the high energy and the long life of the BL. A possible example of high temperature superconducting circular current around the tube of a torus, is a model proposed by Birbrair \cite{Birbrair01}. In his model the shape of BL is not a common sphere but he note that many different forms including the torus are also observed. In this approach we can explain both the high energies (100 kJ) and the small intensity of radiation, for exploding BL. Turner, \cite{Turner94} in order to explain the structure and stability of BL considers a central plasma core surrounded by a cooler intermediate zone in which recombination of most or all of the high-energy ions takes place. Further out, is a zone in which temperatures are low enough for ions to become hydrated. Moreover, near the surface of the ball there is a region in which a thermochemical cooling process can take place. Powell and Finkelstein \cite{Powell69} model assumes that a BL is powered by the electric field which exists between the Earth and cloud base. They suggest that for a BL with typical size and temperature (2000-2500 K) the multiplication of electrons by atomic collisions should be sufficient to sustain the plasma at realistic electric fields.
Muldrew, \cite{Muldrew90}, considers a mathematical model of BL assuming that a solid, positively charged core exists at its center. The large amount of energy occasionally associated with BL is mainly due to the electrostatic energy of the charge on the core. The upper energy limit is determined by the size and strength of the core and this energy can be orders of magnitude greater than the energy which can be confined by atmospheric pressure alone. A pure electron layer and a plasma layer surround the core.  An electromagnetic field is completely trapped by the electron and plasma layers. The electron temperature is sufficiently high that absorption by electron-ion collisions is small, enabling the ball to have a lifetime of seconds or more. Gilman, \cite{Gilman03} suggests a model that consists of highly excited Rydberg atoms with large polarizabilities that bind them together, with cohesion properties that comes from photon exchange forces instead of electron exchange forces. In this model we assume that the density of a BL must be comparable with that of air and it is able to explain the deformability property of some observed BL. Torchigin, \cite{Torchigin05} prefers a radically new approach, where a BL is not composed by material particles but it is a pure optical phenomenon where only an intense light and compressed air interact. In this model a BL is a light bubble which shell is a thin film where the refractive index n is increased as compared with the near space. The shell confines an intense light circulatin within it in all possible directions. This nonlinear optical model can better explain most of the irregular motions and shapes of some BL. The model of Tsintsadze, \cite{Tsintsadze08} is based on a weekly ionized gas in which the electromagnetic radiation can be accumulated through a Bose-Einstein condensation or density inhomogeneity of plasma. This model can explain the observed stability of BL, its motion and deformability; further, it can explain the external conditions for instability and its explosive disappearing. The model presented by Coleman, \cite{Coleman06}, is based on burning atmospheric vortices where combustion is the source of the observed luminosity. This model can explain the complex and irregular motions of BL.
An extensive list of current difficulties which cause the current modelling problems is contained in the review by Turner \cite{Turner02}. For a very detailed list of observational properties and characteristics delineated from the numerous surveys of eyewitness reports and additional physical parameters that have been estimated for BL phenomenon on the basis of statistical analysis performed on the surveys, we refer to the review by Davis, \cite{Davis03}.   
         
We can conclude this short review on the subject of essential properties of ball lightning by quoting the words of Turner, \cite{Turner02}: "Because we do not know how to make (laboratory) long-lived ball lightning or to model what seem (all) the crucial processes, we are forced to use any published material which is potentially related to any ball lightning property if we wish to make progress more rapidly than in the past".

\section{Video-camera observation: a report}

On a cloudy and almost raining summer day of the past year (June 20, 2010) one of the authors (P.V.), during a planned trekking, he was walking along a mountain path near the town Pruno (Stazzema, Lucca - Italy) located at 470 m above sea level. The meteo conditions was very bad on the morning with variable intensity rainfall, interrupted by moments of "quite". There was no wind and temperature was around 15-18 C. At a given point, early in the afternoon, he stops on a little bridge that crosses the Deglio's river that, due to heavy rains even the day before, had an exceptional flow rate. He decide to stay there to catch a short movie of the river upstream and, in particular, to frame a precise area of the river from a distance of about 20 m by using the zoom of his digital video-camera. The movie lasts about six seconds and in this short time interval, within the monochrome viewfinder, he feels something unexplained strange, but he does not give some attention. Only in the final editing of the video, he understands what was strange. In a totally random way in that shot he took over a small ball of orange light, but more white inside the nucleus, moving with irregular motion, with estimated size of a few centimeters. Within those six seconds, the object remained visible for about three seconds, maintaining a constant size and brightness, after which it suddenly vanishes. Halving the speed of the movie, in fact, it is noted that instead of vanishing abruptly the ball accelerates upwards in a diagonal line and then vertically, leaving the camera field of view. At the time of the video recording was not raining and the sky was partially covered. The witness did not receive any special smell, no sound or noise except that of the flowing river.
Was it a BL? The luminous ball recorded in the video looks like to a BL, but we need a detailed image analysis to see if this object can be a good candidate for a true BL.
The reader interested to know more on the video-camera recording can refer to the following Web link: http://fulmineglobulare.xoom.it

\section{Data Analysis}

In this section we report the processes used to eleborate the video and related images. The original video is in a digital format and thus it is possible to perform a digital photometric RGB analysis of the recorded luminous ball object. We recall here that the RGB model is commonly used for the sensing, representation, and display of images in electronic systems.  An RGB-color is a (red, green, blue) vector. Components are here integers between 0 and 255. An RGB-system is in close connection with the consolidated tristimulus color-vision theory of Young - Helmholtz - Maxwell for human. It is based on 3 cones with maximal sensitivity at 564, 534 and 420 nm supplemented with brightness channel of rods. Further, we show a basic dynamical analysis of the ball image in the time interval of its visibility, taking into account the real physical scales of objects as recorded in the video. In order to determine the physical scales, we went in the exact observation site to make accurate measurements of the background objects. The distance between the viewpoint and the area of appearance of the ball was about 17.6 meters, while the comparison between the size of the surrounding rocks and the image in the movie was able to estimate the size of the pixel in the object plane of about 2.2 mm. 

 \subsection{Ball Image analysis}

A photometric RGB analysis of the ball image required the selection of different frames extracted from the original video. In particular, we have identified three frames taken from the original movie corresponding to three different positions of the ball with three different conditions of the background. Further, we performed the analysis of spatial variation of intensity of the luminous "sphere", measured along a diameter, splitting the RGB channels.  Note that although the object shape is near spherical one, in the following frames, due to the video format conversion, the ball images appear slightly vertically elongated.   
  
Figure 1 shows the selected frame in the first position. Note the location of the ball near the edge of the border between the background consisting of water and rock. In order to characterize both the ball and the background, we display in Figure 2 a portion of the frame around the ball and its RGB analysis. In the figure are reported both the pixel scale and the scale on the object plane, appropriately scaled for a direct comparison. We note firstly that the intensity of the emission of the water is higher (about a factor of 2) than that of the rock. Thus, the emission of background where is located the object is decreasing from left to right generating an asymmetric global emission. Secondly, the water emission (on the left) is well RGB characterized, with a predominant B channel, followed by green and red channels. To support this fact, we show in Figure 3 the emission from a water background in a different location (in the middle of the fall) where was present only water. On the other hand, the rock emission is not well RGB characterized, showing only an intermittent dominant red component. Using the above background characterization, and the fact that the object emission is characterized by a strong red dominant component, we have identified the spatial extension of our phenomenon, on the basis of the relative RGB behavior, as shown in figure 2 by vertical bars. The left bar was positioned at the inversion between the B (water) and R (object) components;
whereas the right bar was positioned where the R component is no more dominant. The spatial extension estimated of the ball is: 3.26 cm (\begin{math} \pm 10\% \end{math}). This value is compatible with the accepted size range, as repoted in literature, of a typical dimension for a small BL. Inside the region, we can note the presence of a intense central peak with an extension of about 1.5 cm and two secondary peaks corresponding to a ring feature of the image.

In the following images we reported only the region around the central peak.
            
Figure 4 shows the RGB analysis of frame in Figure 1. The profile of RGB emission curves clearly shows that there is a red dominant component in the light emitted from the ball followed by green and blue ones. The value of R peak is: 218, whereas the base level is: 135. The extension of the main peak central core is: 1.5 cm. 
  
Figures 5,6 show the second selected frame and its RGB analysis. Note that here the position of the ball is now located with the background consisting of rock. The profiles of RGB emissions confirm the characteristics of previous analysis. The value of R peak is: 228, whereas the base level is: 103. This lower value is due to a lower emission from the background (rock). The extension of the main peak central core is: 1.6 cm.

Figures 7,8 show the third selected frame and its RGB analysis. Note that here the position of the ball is located with a background consisting of water. Again the profiles of RGB emissions confirm the main characteristics of previous analyses. The value of R peak is: 212, whereas the base level is: 165. This higher value is due to a higher emission from the background (water). The extension of the main peak central core is: 1.5 cm.

From the last three figures we note that the maximun red emissions are almost the same, whereas the background emission is very different. In order to confirm this characteristic in Figure 9 we show the maximum values of R emissions and the relative background emissions for six different positions of the ball. The different locations are displayed in the inset, and correspond to the complete observed path of the ball covering very different background conditions. We stress here that the R peak values are almost constant whereas the background levels are strongly variable (about a factor of 3). This supports the fact that the ball is not "transparent" in the visible and the emission is constant during the observed time interval.      

From the above RGB analysis of the selected luminous ball images we can draw the following conclusions:

1) The RGB analysis shows that the type of light emission from the ball is not monochromatic; \\
2) There is no saturation of the image; \\
3) From 1) and the absence of blinking of the ball, mainly in the position related to turbulent water, would be excluded, as an explanation of the image, the projection of a laser source at a distance; \\
4) the light emission of the object dominates in the red band in all positions; \\
5) the detected intensity of light is emitted from the source and it is not a reflection from any kind of an external source and is almost constant; \\
6) the peak intensity does not change significantly when changing the emissivity of the background: the object is not "transparent" in the visible; \\
7) estimated size, motion, stability, type of light emission and the environmental high-humidity conditions, would suggest a probable ball lightning. \\

To strengthen the conclusion in 3), after a preliminary characterization of the CCD video-camera response to three different types of lasers at wavelengths: $\lambda=635.6$ nm (red), $\lambda=543$ nm (green) and $\lambda=594$ nm (orange), we projected the spot of red laser in the same area (and weather conditions) of initial observation of the ball. The related RGB analyses clearly showed that the luminous ball object recorded by the video-camera can not in any way be attributed to a diffused and/or a reflected monochromatic light from a laser. A laser image, unlike the ball image, is monochromatic. (For further details on laser tests see the site: http://fulmineglobulare.xoom.it)

In the following dynamical analysis, we will further support the above conclusions based on RGB emission analysis.

\subsection{Ball Dynamics}

We analyzed the motion of the luminous ball recorded in about 2 seconds of its visibility in quasi-stationary conditions, before his sudden demise with great acceleration.
To determine the dynamics of the ball, we initially split the movie into individual frames, the frames were then aligned (stacking) isolating the image of the ball by eliminating the background context. The result is shown in Figure 10. As we can see, from a first inspection, the ball shows a high variability in the motion with quasi-stationary periods alternated with periods of high speed. Of course, we dropped the final frames where the ball accelerates disappearing. Starting from the individual frames, we extracted the coordinates (in pixels) of the "core" of the ball and transformed into the known physical size scale reported above.
Limiting ourselves to the dynamics in the plane perpendicular to the line of sight (transverse components) the calculation of Euclidean distances, given that the interval between two successive frames is 1/25 s, led to a graph of speed shown in Figure 11.
We can easily see continued acceleration and deceleration, with a few intervals at a constant speed. The peak shapes show a dynamics of ball which must be resolved well after the typical scan rate of movie: 1/25 s. Most acceleration occurs at about 1.59 s where the velocity rose from 37.5 cm/s to 180.3 cm/s in 0.04 seconds (acceleration: 3.6 g), and next reduced to 12.5 cm/s in 0.16 seconds. Here no rotation of the ball was detected.
After this analysis of the dynamics of the ball we can safely say that the type of motion and the quantitative estimates of speed, agree well with the data and current models of ball lightning.
To complete the analysis, we also performed a sound analysis: The file of environmental noise sampled at 8kHz has been subjected to a wavelet analysis to highlight the possibility, during the time of appearance of the ball, to detect a noise characteristics of the BL. However, the obtained results clearly showed no evidence of different specific frequencies from those of the ambient noise background.

\section{Comparison with models}

Following the results shown in the preceding sections we now make some brief remarks on the characteristics of ball recorded and some of BL models proposed in the literature. We have already noted that the size (about 4 cm), the light emission (almost constant), motion (quite variable) and duration ($> 3 s$) of the ball are within the parameters allowed and accepted for a BL.
The profiles of the RGB curves are in agreement with the electrochemical model of Turner \cite{Turner94}, \cite{Turner02}, where a hot core plasma is surrounded by shells at lower temperature where ion recombination processes and hydration take place.
The shape of the object is almost spherical one, and then, to explain the observed ball, we can exclude those models that consider more complex geometries.  The observed relatively short duration of the ball does not require the assumption of an external sources of support or the use of electromagnetic models, such as by Ignatovich \cite{Ignatovich06}, to explain a long lifetime of BL. The analysis of ball motion, even when it is seen projected onto a plane perpendicular to the line of observation, highlights both a horizontal component, common to many BL (54\%), but also a significant vertical component (19\%) that only few models try to explain, (Stenhoff, \cite{Stenhoff10}). The nonlinear optical model, by Torchigin et al. \cite{Torchigin05}, appears quite consistent with our results (e.g. observed emissivity, dimension, motion) but it needs further investigations to justify a long lifetime, greater to 3 sec. As reported by Stenhoff \cite{Stenhoff10}, a ball lightning is a relatively low energy phenomenon with maximum energy up to 3 kJ. As reported, a ball visible in daylight is comparable to a 150 W filament lamp, and considering a luminous efficiency of about 20\% it emits about 30 W of power in the visible part of the spectrum (Stenhoff, \cite{Stenhoff10}). In our case, considering that the volume of the observed BL is about $18 ~ cm ^{3}$ we obtain a maximum energy density of about $165 ~ J cm ^{-3}$. In order to estimate the effective power emitted by our ball, we need an energy calibration of the video camera that we would like to perform in a future work allowing a direct comparison with the literature data.  If we consider the light emitted in the visible part of the spectrum, for a yellow-orange ball with $\lambda_{max} = 588 ~ nm$ the blackbody temperature, from the Wien law, would be about 4900 K. Red ball lightning would be cooler and blue ball lightning hotter. However, we have to consider that this is a quite unrealistic estimate based on a pure theoretical relation and it is not clear if the source of luminous energy is thermal or not (Stenhoff, \cite{Stenhoff10}).

\section{Conclusions}

While in principle the existence of ball lightning is generally accepted, the lack of a conclusive, reliable and accurate theory has been partly responsible for some remaining skepticism about their real existence. In fact, most physicists have given a description in terms of plasma physics, but more detailed considerations, based on observations, have given rise to many unexpected problems. This has led to suggest new concepts and interpretations of this phenomenon that have been developed in many common areas of physics and also to assume more or less exotic phenomena such as antimatter, new fundamental particles (dark matter), little black holes (Rabinowitz, \cite{Rabinowitz01}). However, the authors believe that this atmospheric phenomenon can be well described within the physics of plasma, the electrochemical processes, nonlinear optics and electromagnetic fields. Since at present the only way we can define the properties of ball lightning is through the direct accounts of observers, it is crucial to consider the reliability and accuracy of reports. This fact introduces a certain degree of subjectivity that depends on the experience and knowledge of the observer. In this paper, the authors aim to contribute to the wide collection of observational reports through the objective analysis of a recorded video in the daytime, which strongly suggests the presence of a small moving ball lightning, and that provides to specialists in the field new material for further detailed studies. The main goal here was to provide some useful information, not subject to personal interpretation, to improve some aspects of the proposed models and to better explain the rare phenomenon of ball lightning. Following this viewpoint future work could be useful to give: 1) a more precise estimation of the temperature inside the ball by a more deep analysis of the spectral composition of emitted light and a comparison with a blackbody emission; 2) a more precise estimation of the power of emitted light; 3) a deep investigation of the spatial structure of the ball (e.g. multiple rings); 4) a more advanced dynamic analysis of the ball, e.g. with a 3D motion reconstruction using proper image postprocessing algorithms.

\newpage

\begin{figure}
\resizebox{\hsize}{!}{\includegraphics{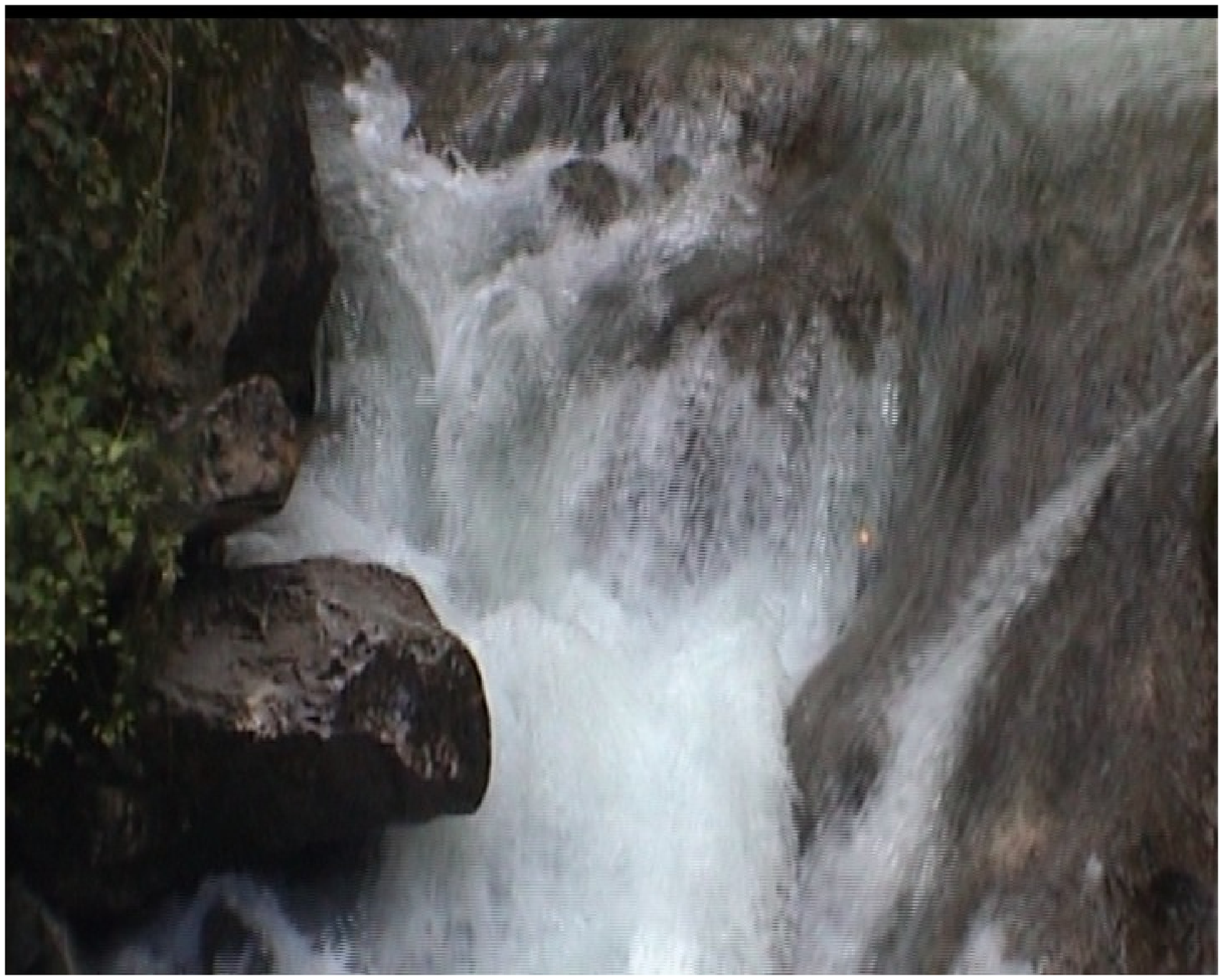}}
 \caption{First selected frame with a luminous ball.}
 \label{fig1}
\end{figure}

\begin{figure}
\resizebox{\hsize}{!}{\includegraphics{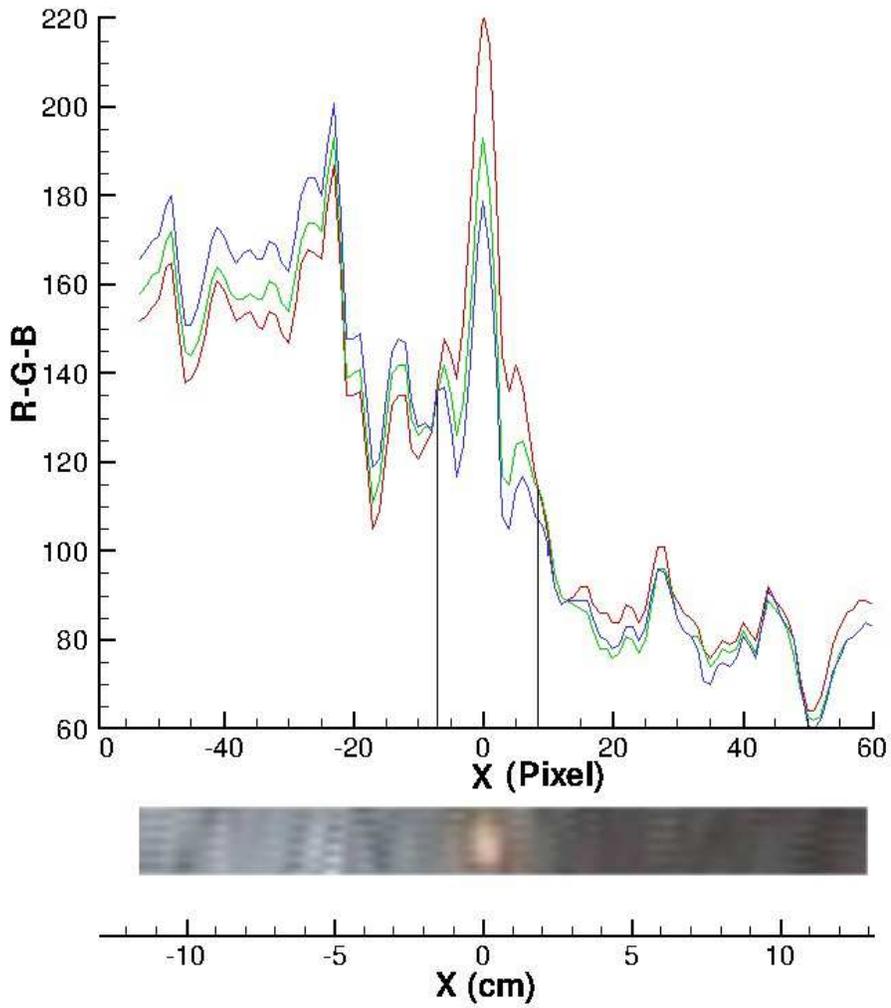}}
 \caption{RGB emission for an extended spatial region near the object (1px=2.2mm).}
 \label{fig2}
\end{figure}

\begin{figure}
\resizebox{\hsize}{!}{\includegraphics{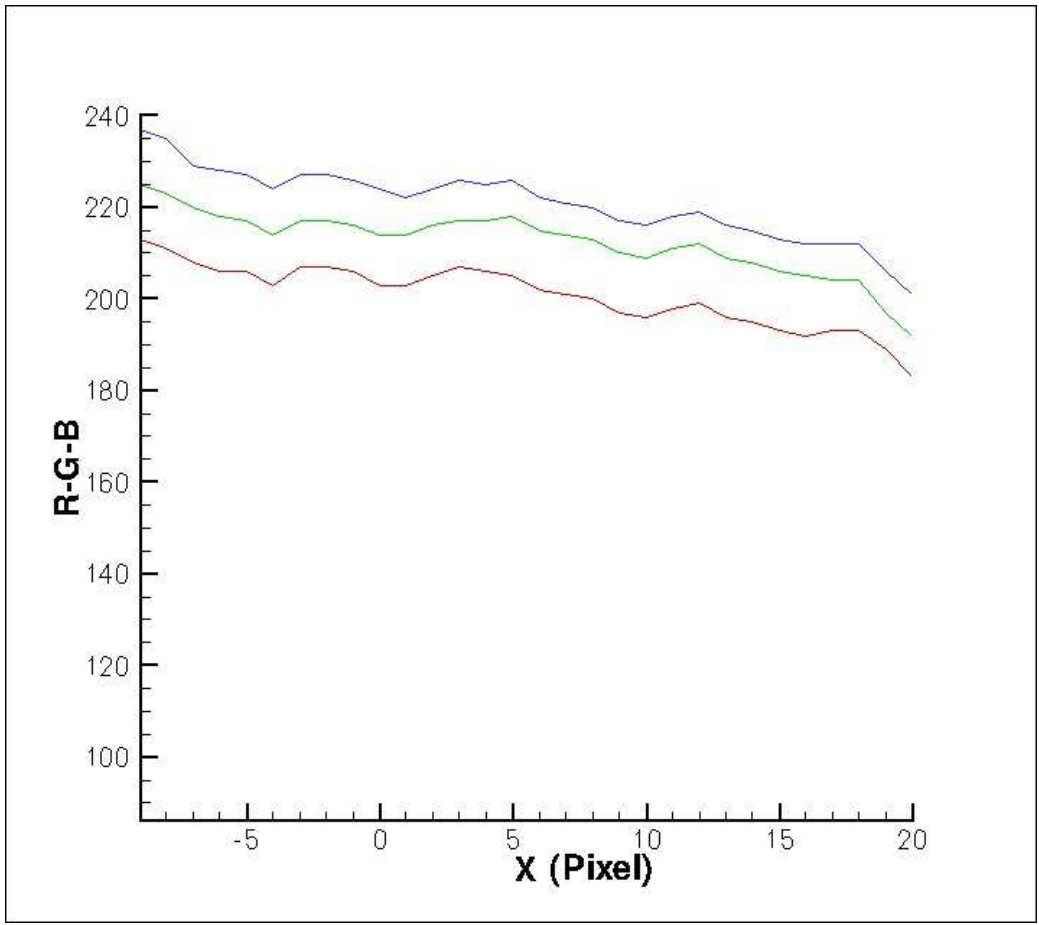}}
 \caption{RGB emission from water around the object (1px=2.2mm).}
 \label{fig3}
 
\end{figure}
\begin{figure}
\resizebox{\hsize}{!}{\includegraphics{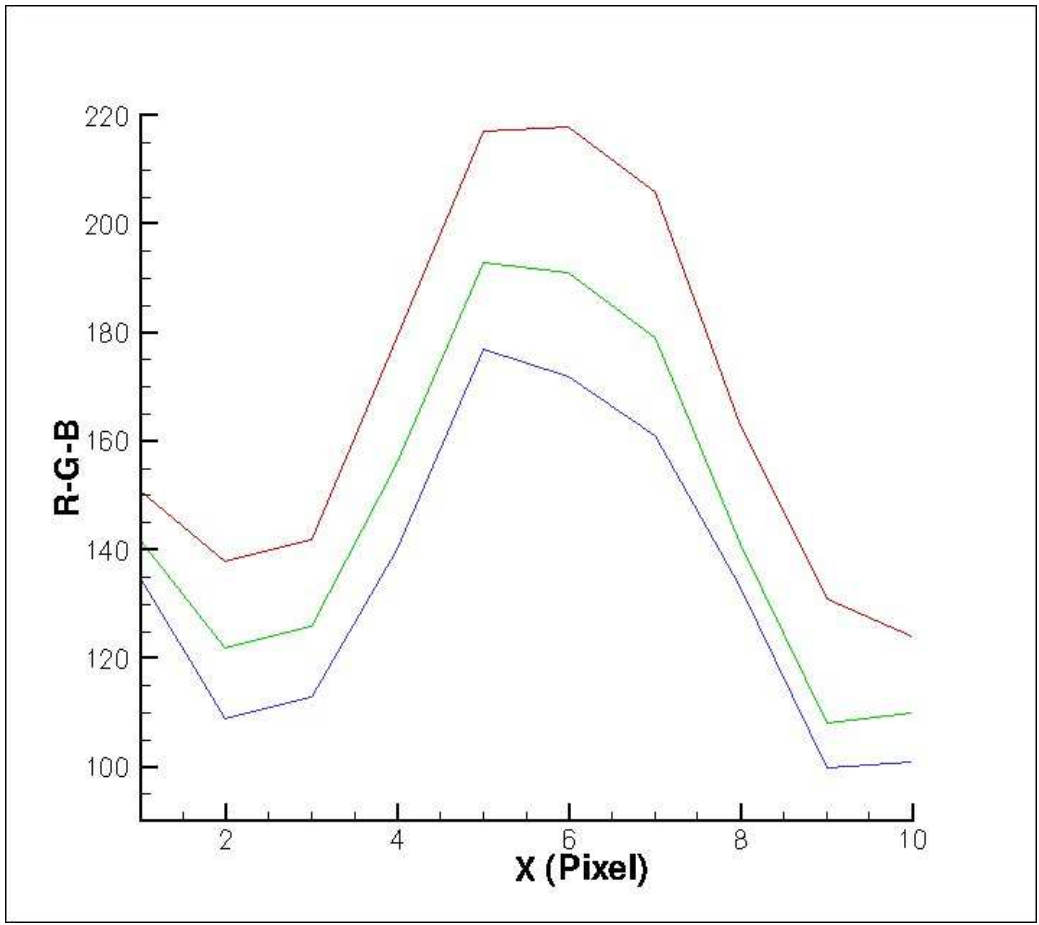}}
 \caption{RGB analysis of first selected frame (1px=2.2mm).}
 \label{fig4}
\end{figure}

\begin{figure}
\resizebox{\hsize}{!}{\includegraphics{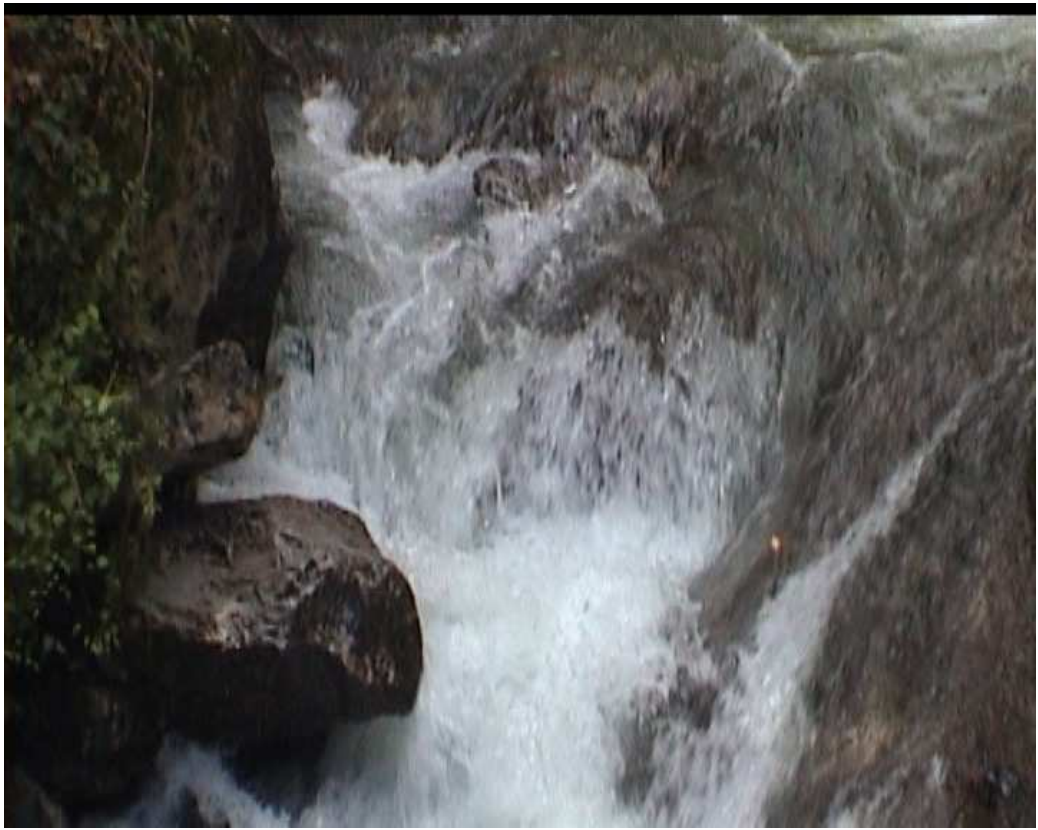}}
 \caption{Second selected frame with a luminous ball.}
 \label{fig5}
\end{figure}

\begin{figure}
\resizebox{\hsize}{!}{\includegraphics{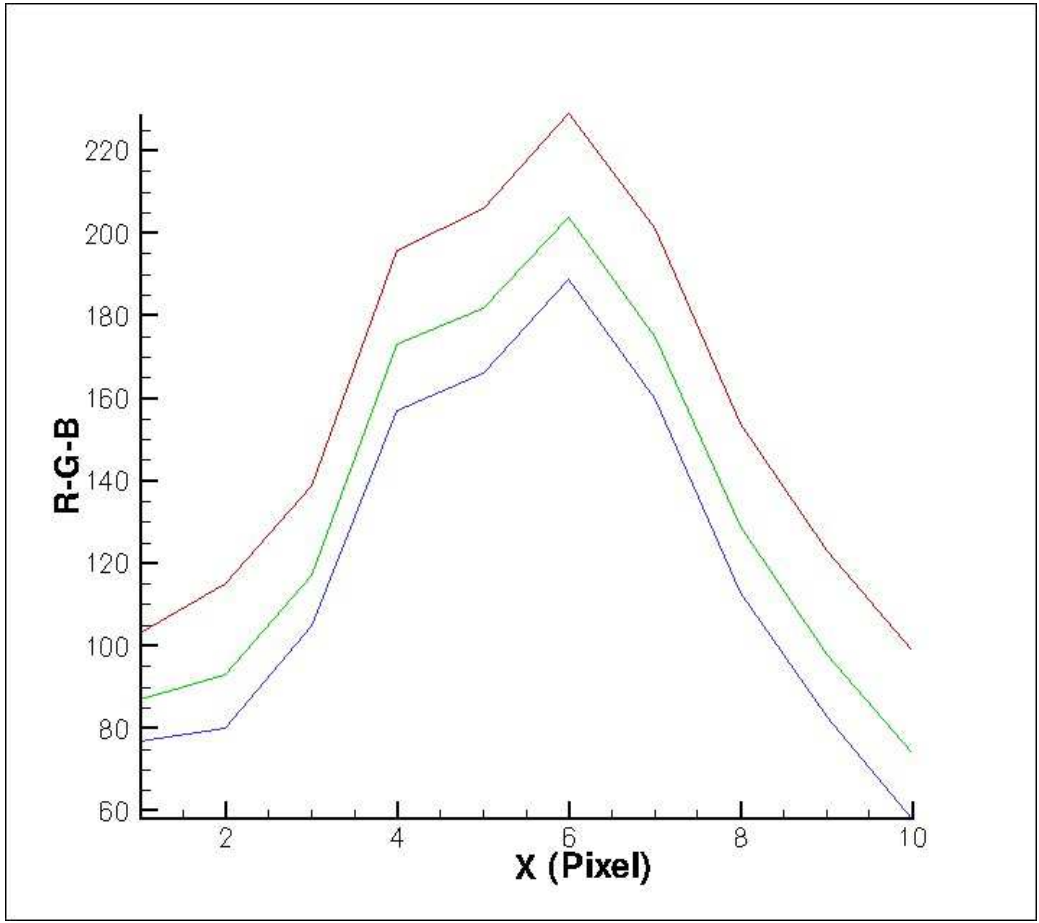}}
 \caption{RGB analysis of second selected frame (1px=2.2mm).}
 \label{fig6}
\end{figure}

\begin{figure}
\resizebox{\hsize}{!}{\includegraphics{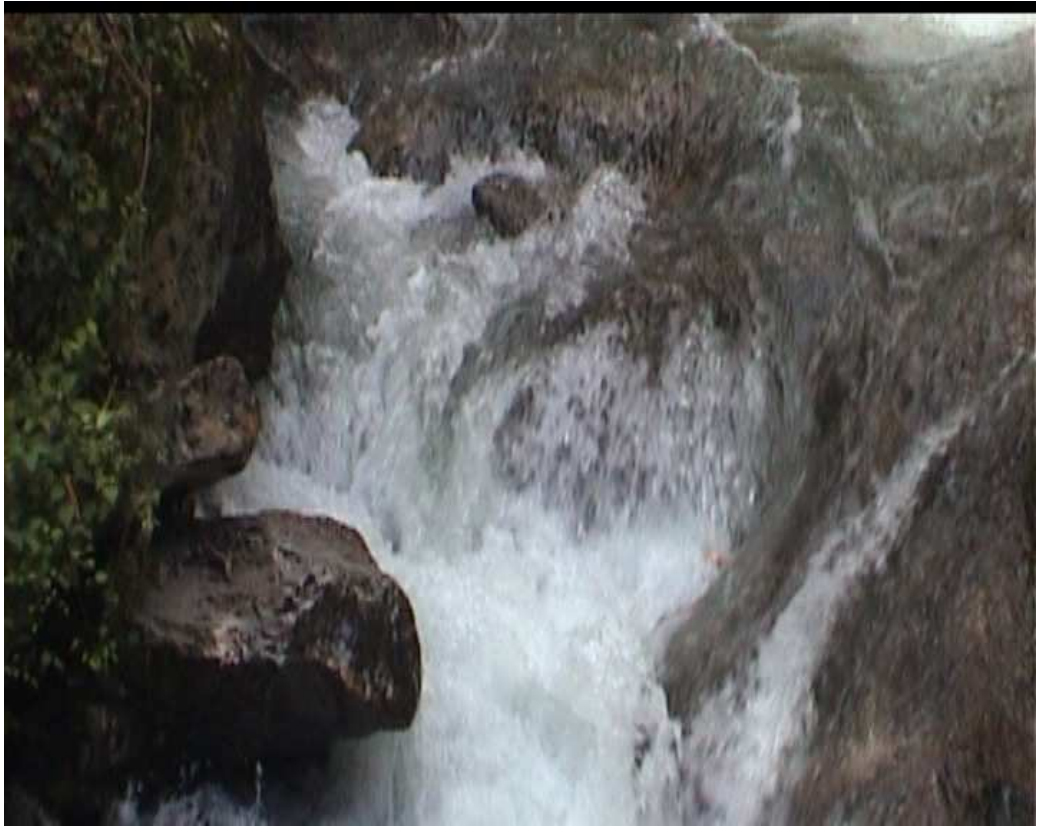}}
 \caption{Third selected frame with a luminous ball.}
 \label{fig7}
\end{figure}

\begin{figure}
\resizebox{\hsize}{!}{\includegraphics{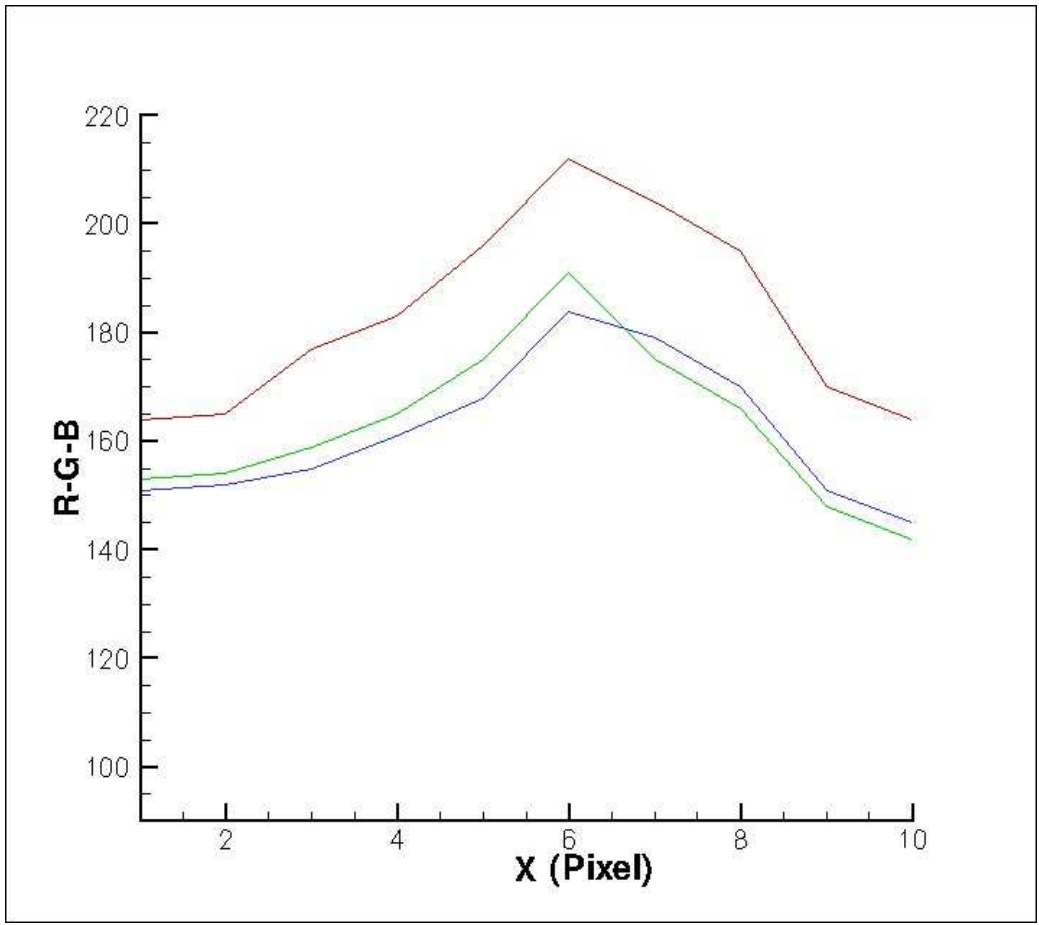}}
 \caption{RGB analysis of third selected frame (1px=2.2mm).}
 \label{fig8}
\end{figure}

\begin{figure}
\resizebox{\hsize}{!}{\includegraphics{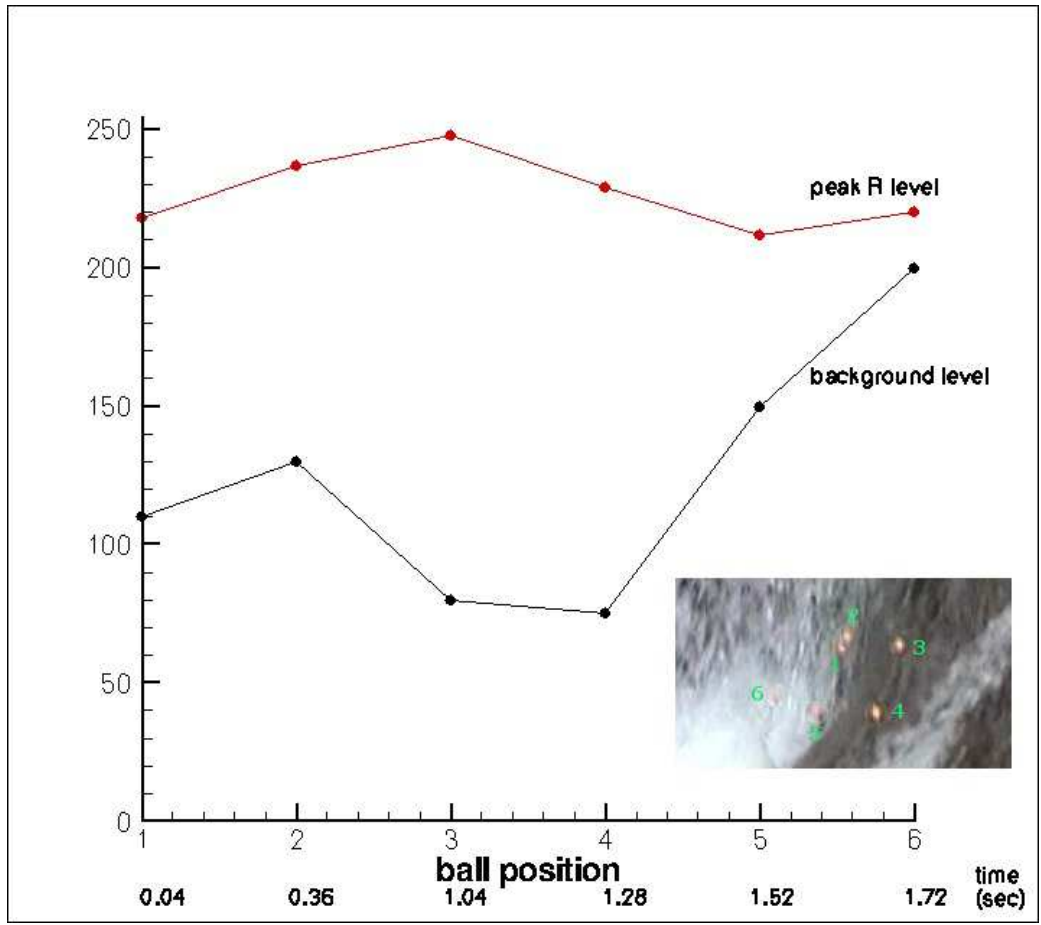}}
 \caption{Peak R values for six different positions of the object corresponding to different background levels (see inset).}
 \label{fig9}
\end{figure}

\begin{figure}
\resizebox{\hsize}{!}{\includegraphics{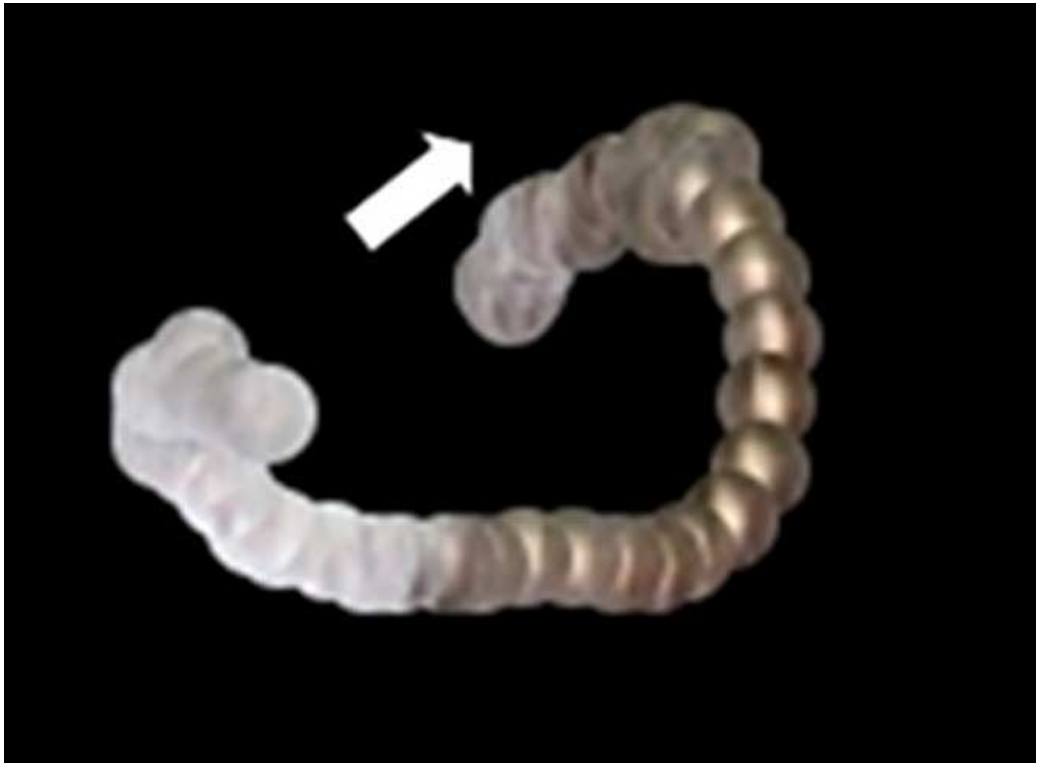}}
 \caption{Combined snapshots of ball positions}
 \label{fig10}
\end{figure}

\begin{figure}
\resizebox{\hsize}{!}{\includegraphics{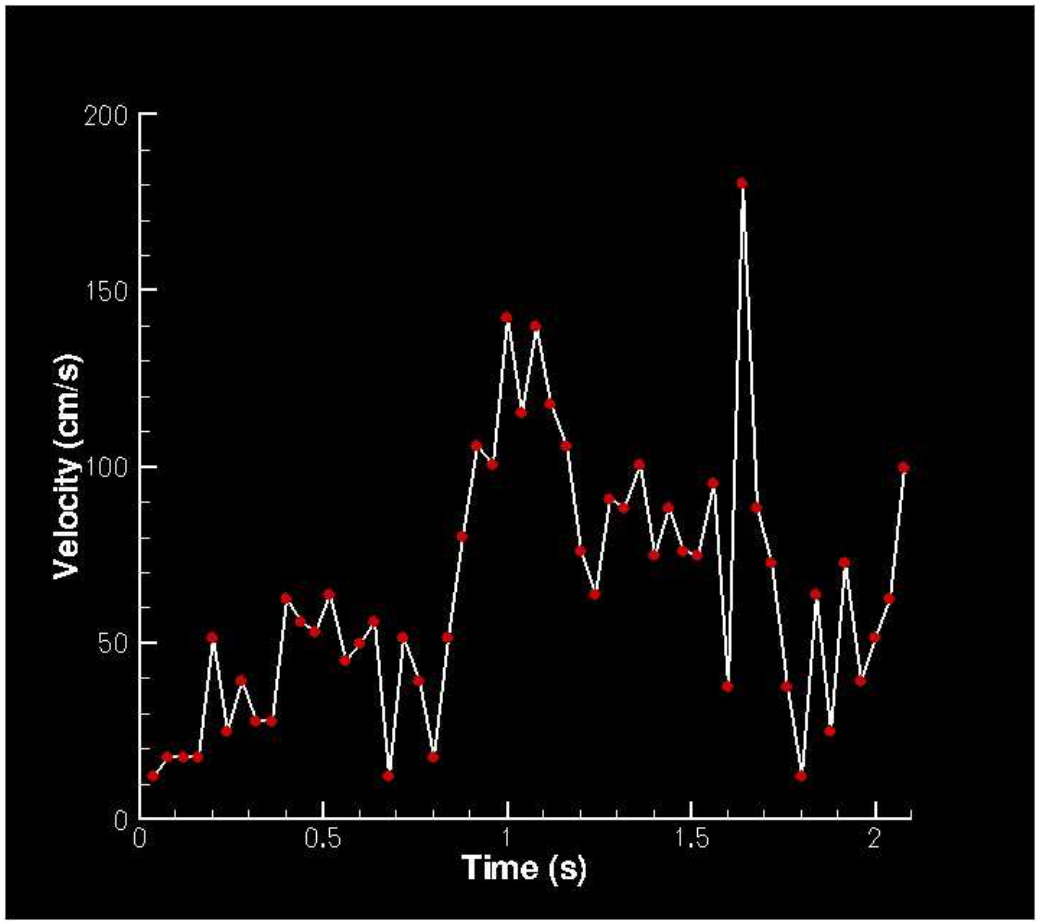}}
 \caption{Graph of ball velocities derived from individual frames}
 \label{fig11}
\end{figure}

\end{document}